# Super-geometric electron focusing on the hexagonal Fermi surface of PdCoO$_2$

Maja D. Bachmann[1,2,†], Aaron L. Sharpe[3,4,†], Arthur W. Barnard[5], Carsten Putzke[1,6], Markus König[1], Seunghyun Khim[1], David Goldhaber-Gordon[4,5], Andrew P. Mackenzie[1,2] and Philip J.W. Moll[1,5,*]

[1]Max-Planck-Institute for Chemical Physics of Solids, 01187 Dresden, Germany
[2]School of Physics and Astronomy, University of St. Andrews, St. Andrews KY16 9SS, UK
[3]Department of Applied Physics, Stanford University, Stanford, California 94305, USA
[4]SLAC National Accelerator Laboratory, Menlo Park, California, 94025, USA
[5]Department of Physics, Stanford University, Stanford, California 94305, USA
[6]Institute of Materials, École Polytechnique Fédéral de Lausanne, 1015 Lausanne, Switzerland
[*]To whom correspondence should be addressed: bachmann@cpfs.mpg.de, philip.moll@epfl.ch
[†]These authors contributed equally to this work.

Geometric electron optics may be implemented in solid state when transport is ballistic on the length scale of a device. Currently, this is realized mainly in 2D materials characterized by circular Fermi surfaces. Here we demonstrate that the nearly perfectly hexagonal Fermi surface of PdCoO$_2$ gives rise to highly directional ballistic transport. We probe this directional ballistic regime in a single crystal of PdCoO$_2$ by use of focused ion beam (FIB) micro-machining, defining crystalline ballistic circuits with features as small as 250nm. The peculiar hexagonal Fermi surface naturally leads to electron self-focusing effects in a magnetic field, well below the geometric limit associated with a circular Fermi surface. This super-geometric focusing can be quantitatively predicted for arbitrary device geometry, based on the hexagonal cyclotron orbits appearing in this material. These results suggest a novel class of ballistic electronic devices exploiting the unique transport characteristics of strongly faceted Fermi surfaces.

Electronic conduction in metals is typically well captured by Ohm's law as frequent collisions of the electrons lead to diffusive motion and locally-defined conductivity. An essential prerequisite to this common transport regime is a momentum relaxing mean-free-path, **λ**, that is much smaller than the size of the conductor. In extremely clean metals, however, **λ** may exceed the size of the sample, and a ballistic description of charge transport becomes appropriate. The diffusive motion of Ohm's law is replaced by ballistic trajectories, akin to the motion of billiard balls. In such situations, geometric electron-beam optics should be achievable in solid-state devices. Significant progress has been made in two-dimensional electron systems (2DESs), where essential elements of electron optics (familiar from a free-space context) have been demonstrated in high-purity semiconducting heterostructures and graphene. These include collimated electron sources[1,2], lenses[3–5], waveguides[6,7], beam splitters[8], refractive[9,10] and reflective[11] elements. Electronic solid-state devices based on high carrier density metals differ from free-space electron beam applications as they operate in a quantum regime of a Fermi gas at temperatures far below their Fermi energy. Therefore, the accessible electron states for conduction are locked to the Fermi energy and the Fermi momentum, unlike the free space electronic beams where all energies and momenta are accessible.

To date, such device concepts rely on straight-line electron trajectory segments or, in magnetic field, circular orbits associated with small, isotropic Fermi surfaces (FS) as in graphene or 2DESs. In principle, unlike the free electron case, a solid offers the opportunity to engineer dispersion relations E(**k**) via Bragg scattering off the lattice. By tuning the hopping integrals, the shape of the FS can be made strongly non-circular defining preferred directions of electronic motion in the solid. We report here a striking directionality of ballistic electron motion in the material $PdCoO_2$, arising from its almost perfectly hexagonal FS which defines three preferred directions of motion.

The metallic delafossite $PdCoO_2$ is an extraordinarily clean conductive oxide which exhibits a mean-free-path of ~20µm at low temperatures[12], rendering the delafossite metals class the most conductive oxides known[13–16]. The quasi-2D crystal consists of highly conductive, triangular coordinated palladium sheets separated by layers of insulating $CoO_2$ octahedra. Only a single, half-filled band crosses the Fermi level, resulting in a cylindrical FS consisting of a nearly perfectly hexagonal cross-section weakly warped along $k_z$[17,18]. Unusual transport characteristics, such as strong momentum-conserving scattering processes, which have been argued to lead to hydrodynamic transport[19], have recently attracted attention. Here we focus on the exotic ballistic regime of $PdCoO_2$ at low temperature (2K) and in a range of applied out of plane magnetic fields (B<=14T), arising from transverse electron focusing (TEF) in combination with the hexagonal Fermi surface.

A typical geometry for a TEF experiment is sketched in Fig. 1a. Two narrow contacts resembling nozzles connecting along the same crystal edge and two far away large electrodes form the electric connections to the device. A uniform magnetic field B is applied perpendicular to the surface, causing electrons to follow cyclotron orbits. If the nozzle separation L is smaller than or comparable to the shortest microscopic mean free path in the problem, one expects undisturbed cyclotron motion between the nozzles. Similar to the circular trajectories of electrons in free space, electrons in a lattice move on orbits defined by the geometry of their FS. When an integer multiple n of the cyclotron diameter matches the contact spacing L, an excess of electrons arrives at the distant contact, leading to an increase of the chemical potential there. As the cyclotron radius $r_c = \hbar k_F/eB$, is inversely proportional to the magnetic field B, and electrons back reflect when they hit the crystal edge, a linearly spaced train of peaks in measured potential can be observed at fields $B_n = \frac{2n\,\hbar k_F}{eL}$, where $\hbar$ is the reduced Planck constant, $k_F$ is the Fermi vector and e is the electronic charge. These correspond to trajectories with n-1 bounces. First observed and studied in elemental (semi-)metals (Bi[20], Sb[21], W[22], Cu[22], Ag[23], Zn[24], Al[25]), the TEF effect was further investigated in 2DESs[11] and graphene[26,27] and has recently been employed to spatially separate and detect electron spins in a spin-orbit coupled system[28].

In $PdCoO_2$ the TEF signal strongly deviates from that expected for free electrons due to the hexagonal shape of the FS. In the presence of a magnetic field, the charge carriers revolve around the FS in k-space in a plane perpendicular to the applied field. In real-space, the shape of the cyclotron motion is given by a 90-degree rotation of the k-space orbit. In materials with circular FSs, like graphene or 2DESs, the electron trajectories are circular and their velocity distribution is isotropic in real space (see Fig. 1b). However, since the group velocity is locally perpendicular to the FS, strongly faceted FS such as in $PdCoO_2$ will have a macroscopic proportion of states at the Fermi energy all moving in the same direction. This leads to a highly anisotropic velocity distribution in the palladium planes with 3 preferred directions of motion.

On a circular FS, a simple geometric model can be used to understand the shape of the TEF peak. An electron injected at an angle $\theta$ away from normal incidence will be focused at a distance $x = 2r_c \cos\theta$ away from the nozzle. Assuming isotropic angular distribution of electron injection, we can find the distribution of distances from the injection nozzle at which electrons return to the edge, by calculating the classical probability density function, which is given by $n(x) = \frac{2}{\pi}\frac{1}{\sqrt{(2r_c)^2-x^2}}$ (see methods for derivation). The divergence at $x = 2r_c$ describes the focusing effect on a circular FS; even when electrons are injected evenly in all directions, those entering the device under a small angle $\theta$ will all be focused onto nearly the same spot by a magnetic field (purple shaded region in Fig. 1c). This occurs due to the presence of a well-defined Fermi surface in the metal and is hard to achieve in free-space electron beams as it requires monochromatic electrons.

On the hexagonal FS of PdCoO$_2$ however, one expects large flat sheets, along which electrons are naturally focused onto the same point (orange shaded region in Fig. 1c). This intuition of an enhanced "super-geometric focusing" compared to that associated with circular FS materials is corroborated by numerical TEF simulations on the experimentally determined hexagonal FS of PdCoO$_2$ (Fig. 1d, see methods for details on simulations).

To test this prediction experimentally, we have fabricated TEF devices from as-grown single crystals of PdCoO$_2$, which grow as ultra clean single crystals without the need for any further purification. The synthesis is described in the methods section and elsewhere[29]. The crystals grow as hexagonal platelets (~10-20μm thick, and several 100μm in lateral dimensions), with the growth edges of the crystal oriented 90 degrees away from the crystal axes (Fig.2a).

Critical for the observation of TEF as outlined above is the use of narrow injection nozzles. In two-dimensional systems, point-like contacts can be easily defined lithographically. Here we employ a focused ion beam (FIB) based technique to fabricate narrow nozzles of diameters as small as 250nm ($\ll \lambda$, L). The details of FIB micro-machining are described in the methods section and can be found elsewhere[30]. As a mask-less technique capable of etching materials in three dimensions, FIB machining offers a unique way to carve a well-defined micro-sample out of a larger crystal. Given the high degree of control over the material on the sub-μm scale, this approach may be a viable route toward a more quantitative Fermiology based on TEF on even the smallest metallic samples[27].

In our typical PdCoO$_2$ devices, two sets of nozzles (Fig.2c, bottom edge: 1-8 and left side: A-D) connect to a central square area and are arranged perpendicular to each other. The orientation of the FS and the Brillouin zone are overlaid in the middle panel. Due to the six-fold rotational symmetry of the FS, these two sets of nozzles will probe TEF directed along different parts of the FS. The bottom panel shows an enlarged view of the nozzles, which are ~250nm wide and are separated by 1μm each. The temperature dependence of the non-local voltages (measurement configuration Fig. 2c; data Fig. 2d) observed in our experiment is reproduced quantitatively by solving the Laplace equation numerically in our geometry using bulk resistivity values[16]. This diffusive picture describes the transport across the entire device well as it is much larger than the mean-free-path at any temperature. The quantitative agreement suggests the FIB fabricated crystalline circuit matches the designed geometry, and the electronic properties over most of the structure are not strongly perturbed by the FIB process.

For a hexagonal FS, the TEF strongly depends on the crystallographic direction (Fig. 3), because the group velocity of a k-state is always perpendicular to the FS. When the electrons are ejected from a nozzle onto a flat side of the FS hexagon, they are collimated into three main directions, hence we term this configuration the '3 beam direction'. If the nozzles are rotated by 90 degrees with respect to the underlying FS, the electrons then travel in a '2 beam direction' configuration. With an applied out-of-plane magnetic field the electrons then undergo hexagonal cyclotron motion, yet with different initial conditions. Accordingly, the experimental TEF peaks from nozzle pairs of the same separation occur at different fields and differ in shape between the two crystal directions, as seen in Fig. 3b. At negative fields, there is a small, diffusive magneto-resistive component. Since there is only one type of charge carrier (electrons) present in this system, no focusing peaks are observed at negative fields. At small positive fields, where the cyclotron diameter is larger than the separation of the nozzles, a voltage inversion is observed: more electrons reach the large contact ($V_{com}$ in Fig. 2c) than the nozzle. Once the cyclotron diameter equals the distance between the nozzles a voltage peak is detected. As expected for the B-linear period of TEF, the second peak of the 4μm spaced nozzles coincides with the first peak of the 2μm spaced nozzles. The Fermi surface is encoded in the peak shape leading to a significant enhancement of the focusing in the 3-beam compared to the 2-beam configuration. While in the 3-beam direction a large number of electrons are focused into a sharp single peak, the 2-beam direction displays a broad shoulder followed by a peak of reduced amplitude. Simple geometric arguments show that for an ideal hexagon, the three-beam direction would exhibit a divergent super-geometric focusing, while the 2-beam direction would not focus at all (see methods and sketch in Fig. 3a). The focusing peak in the 2-beam configuration as well as the rounding of the peak in the 3-beam data arise from the deviations the real Fermi surface from the sharp corners of the mathematical hexagon. Monte Carlo based simulations using a tight binding approximation of the Fermi surface[12] based on ARPES data[17] (see methods) qualitatively reproduce the observed peaks and their fine structure (Fig. 3B).

TEF probes semi-classical trajectories of ballistic electrons, and in the absence of scattering would lead to focusing over arbitrary distances. In real crystals, scattering is always present, limiting the range over which focusing can be observed. Accordingly, the height of the first TEF peak shrinks as the nozzle distance is increased (Fig. 3c). It is intuitively clear that the TEF signal will be strongly suppressed for nozzles between which a typical ballistic path would be longer than a mean-free-path. A more rigorous calculation, comparing to the observed exponential decay of focusing signal with nozzle separation, allows us to extract from the data a mean free path on the order of **λ** ≈ 15μm (see methods). This value estimated from TEF is in good agreement with previous studies estimating **λ** from transport using a simple Drude model[19]. As the temperature is increased, the normalized amplitude of the primary peak stays roughly constant up to 20 K (Fig 3d). This is consistent with a roughly constant – and long – momentum-relaxing mean-free-path in this temperature region, as reflected by the temperature independent resistivity observed self-consistently in the devices (Fig. 2d) and in measurements on macroscopic crystals[12]. From analysis of a previous flow experiment[19], a momentum-conserving mean free path of approximately 2 µm was deduced for $PdCoO_2$ below 20 K. The possible effects of this were not included in the models used for the simulations presented in Fig. 3B or Fig. S7. Although the simulations clearly capture the main features of our observations very well, there are differences of detail, generally seen in the simulations containing sharper features than the experimental data; it is possible that these differences would be reconciled by including momentum-conserving scattering in a more complete analysis. Above 20K, the focusing peaks gradually decrease until they cease to exist around 70K, presumably primarily due to the reduction of the momentum-relaxing mean free path due to Umklapp electron-electron and electron-phonon scattering.

The first peak $B_1$ plays a special role in the TEF geometry as it describes ballistic paths between both nozzles without any interaction with the device boundary. All other peaks at $B_n$, n>1, correspond to paths which include (n-1) scattering events on the sidewall between the nozzles. In analysis in the literature [refs], TEF has been used as a kind of surface spectroscopy, based on the logic that upon impact with the sidewalls of the device, the electrons may be specularly scattered, thus conserving their momentum component parallel to the surface, or diffusely scattered, leading to a random continuation of the trajectory after the impact. The specularity of the surface, p, denotes the probability of specular reflection. Conventionally, the amplitude ratios between subsequent peaks, $q_n = \frac{A_n}{A_{n+1}}$, are used to estimate p, as each subsequent peak corresponds to a trajectory that differs by one additional surface impact[31]. However, even in a simplified analysis adopting the above assumptions, in the super-geometric focusing configuration the point-spread-function is so sharply peaked at $2r_c$ (see Fig. 1d), that electrons will be statistically refocused onto the next peak, regardless of the specularity. In agreement, we observe peaks up to n=8 in the devices, yielding a *q*-value of roughly 0.6 (Fig. 4). Numerical simulations corroborate this finding, showing that even barely specular surfaces (p=0.1) are compatible with a *q*-value around 0.4 (Fig. 4b).

Our results demonstrate the feasibility of novel electronic devices operating in the ballistic limit exploiting strong deviations from circular Fermi surfaces. This additional avenue of control will enable novel types of functionality. For example, selectively aligning parts of a $PdCoO_2$ crystalline circuit along the 3-beam or the 2-beam direction will completely alter the ballistic response of the device, despite it being chemically and structurally homogeneous. Intriguingly, $PdCoO_2$ is an extremely conductive ballistic metal, and therefore automatically incorporates low dissipation, a key prerequisite for high power and high frequency applications that is unlikely to be achievable in low carrier density devices based on graphene and 2DES. Promising recent thin film results[32] may indicate a pathway towards larger scale fabrication of such devices.

The almost perfectly hexagonal FS shape of $PdCoO_2$ arises from accidental fine tuning of hopping parameters in its band dispersion. In general, super-geometric focusing is a generic property of materials with large parallel sections on their FS. Such flat areas on Fermi surfaces can be engineered in 2D materials where arbitrary control over the chemical potential and sometimes even the band structure is possible via gating, such as in bilayer graphene[26] or in graphene-based moiré superlattices[27]. Alternatively, flat Fermi surface sections are not rare in bulk crystals such as $PdCoO_2$, and future material science efforts may uncover ballistic behavior in other ultra-clean metals. Our FIB based approach here showcases a viable route towards the investigation of ballistic behavior in challenging materials, where the crystal size or chemical composition may impede traditional lithography-based methods to fabricate ballistic devices on the sub-µm scale.

**Acknowledgements:** A.L.S would like to thank Edwin Huang for helpful discussions and Tom Devereaux for the letting us use his groups cluster. The project was supported by the Max-Planck-Society and funded by the Deutsche Forschungsgemeinschaft (DFG, German Research Foundation) – MO 3077/1-1. M.D.B. acknowledges studentship funding from EPSRC under grant no. EP/I007002/1. A.L.S. acknowledges support from a Ford Foundation Predoctoral Fellowship and a National Science Foundation Graduate Research Fellowship.

Computational work was performed on the Sherlock cluster at Stanford University and on resources of the National Energy Research Scientific Computing Center, supported by DOE under contract DE_AC02-05CH11231. Data supporting this manuscript are stored on the Sherlock cluster at Stanford University and are available from the corresponding author upon request. Source code for the simulations can be found at https://github.com/dgglab.

**Author contributions:** M.D.B, C.P., M.K., P.J.W.M. fabricated the microstructures, and M.D.B.,

# Figure 1: Transverse electron focusing on a circular vs a nearly perfectly hexagonal Fermi surface

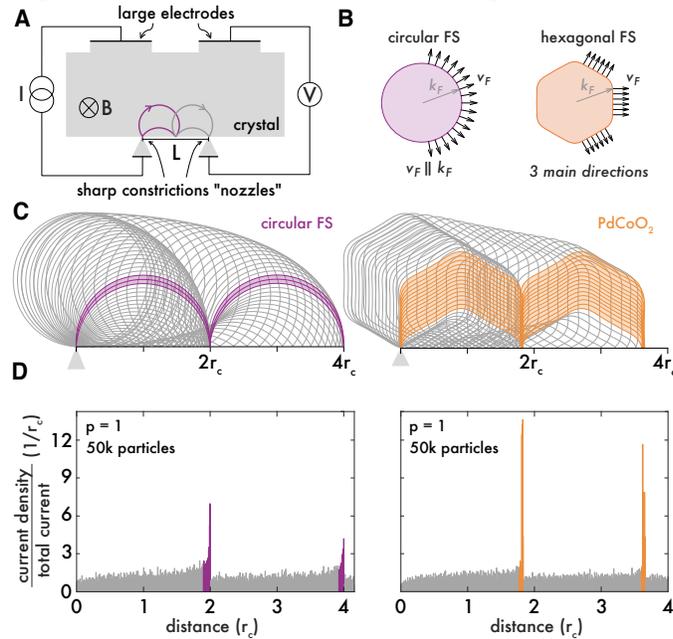

A) Experimental schematic for transverse electron focusing (TEF). Injected through a narrow nozzle, electrons in a 2D system follow an in-plane cyclotron motion when subject to an out-of-plane magnetic field. When a second, receiver nozzle is located at a distance L away from the injection nozzle, the focusing condition is met when the nozzle separation L corresponds to an integer multiple of the cyclotron diameter ($L = n\, 2r_c$). This leads to a focusing spectrum with characteristic voltage peaks at field values corresponding to $B_n = \frac{2n\hbar k_F}{eL}$, where $n$ is a positive integer.

B) Directional restriction of the Fermi velocity $v_F$ on a nearly hexagonal FS in contrast to a circular FS. In the latter case, $v_F$ is always parallel to $k_F$ and can take any direction in real space. On a hexagonal FS however, the large flat sections lead to a collimation of electrons into only 3 main directions.

C) Semi-classical trajectories of electrons injected isotropically at the origin for a circular (left) and nearly hexagonal FS (right). The trajectories contained in the shaded region give rise to peak in the focusing spectrum in the panels below.

D) Simulated focusing spectra for a circular (left) and hexagonal FS (right) assuming completely specular boundaries (p=1). The flat sides of the hexagonal FS lead to a significant increase of the focusing peak height, purely due its geometrical shape.

Figure 2: Focused Ion Beam defined TEF device, measurement scheme and characterization

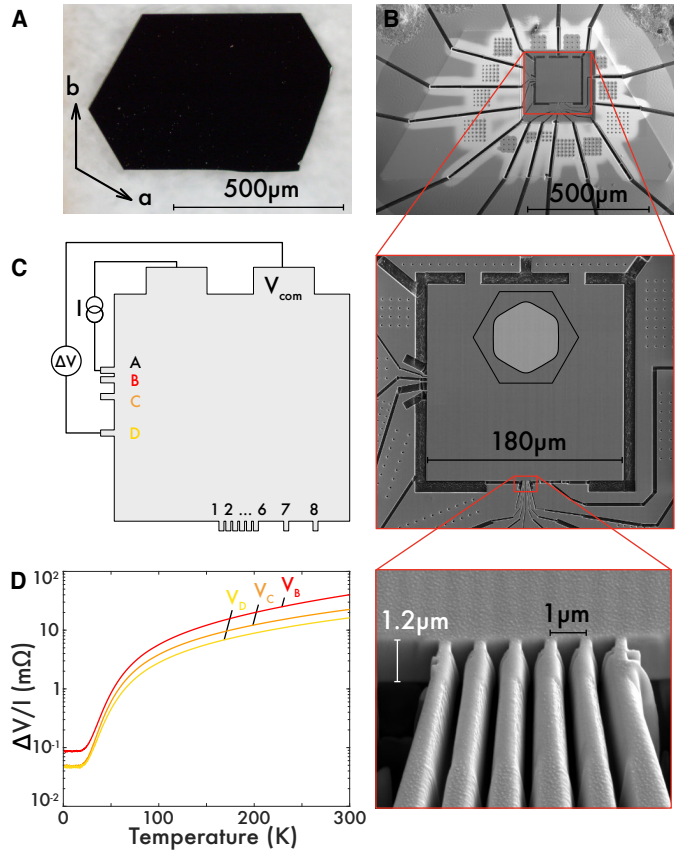

A) Optical microscope image of a single-crystal of PdCoO$_2$ with a thickness of ~20µm. The crystal axes are rotated 90 degrees away from the natural growth edges.

B) Top: Scanning electron micrograph of a FIB defined TEF device. The crystal has been top-contacted with gold and structured into a multi-terminal transport device. Middle: Magnification of the central region. By top irradiation with a gallium ion beam, the crystal has been locally thinned down to ~1µm. Two sets of nozzles, oriented 90 degrees with respect to each other probe TEF along the corner and the flat sides of the hexagonal FS. Bottom: Side view on to the lower set of nozzles, which are separated by 1µm and are about 250nm wide. The long constrictions leading towards the nozzles act both as flexures to reduce fractures and as collimators.

C) Experimental setup for probing directional dependent TEF. In all measurements, the current is sourced between the top left large electrode and a nozzle, and the voltage is measured between the top right electrode (V$_{com}$) and a second nozzle.

D) Non-local voltage signal divided by the sourced current as a function of temperature. The measured voltages are in good agreement with the solutions to the Laplace equation solved with finite element simulations assuming in-plane resistivities of $\rho_{300K}$=3µΩcm and $\rho_{2K}$=8nΩcm.

# Figure 3: Experimental results and ballistic simulations

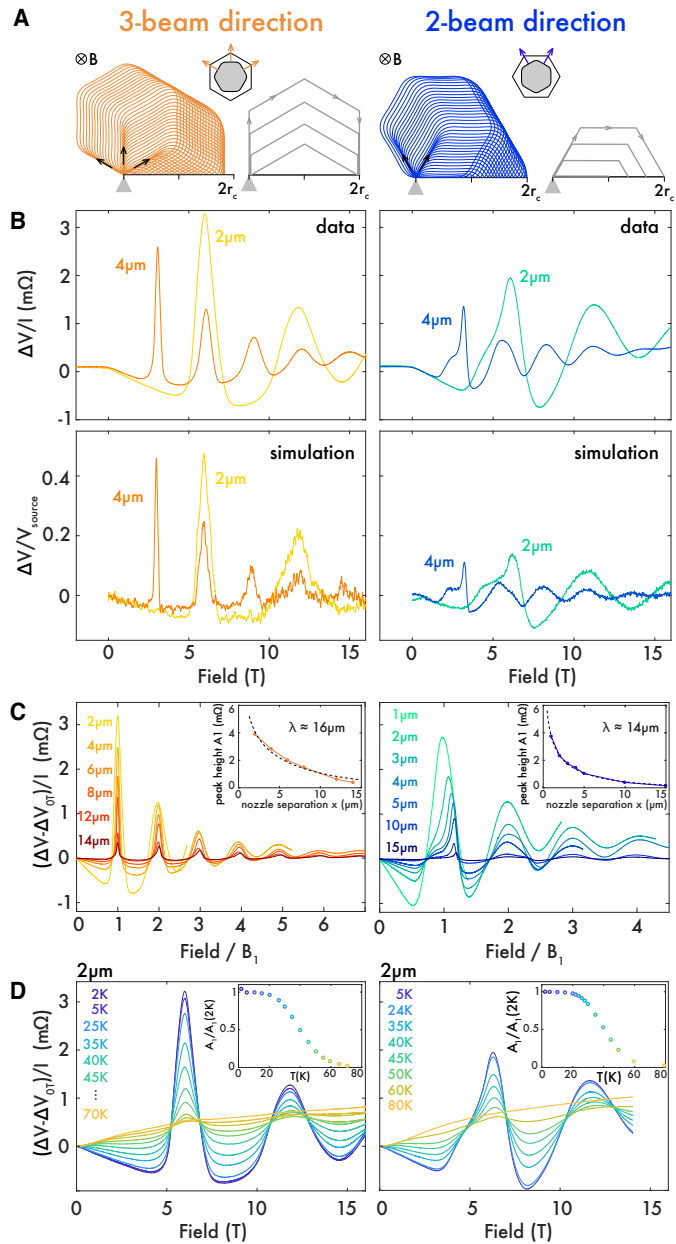

A) Comparison of the TEF geometry for nozzles (grey triangles) cut parallel to the edge of the FS (left) and the corner of the FS (right). In the former case electrons are emitted (grey triangle) predominantly along 3 directions, in contrast to the 30-degree rotated case where only 2 electron jets are formed. The caustics are sketched for perfectly hexagons and while the 3-beam direction shows ideal super-geometric focusing conditions, focusing is absent along the 2-beam direction for a hexagon with sharp corners.

B) Top row: Measured voltages $\Delta V$ divided by the applied current I=3 mA as a function of magnetic field for nozzle pairs separated by 2 μm and 4 μm along the 3-beam (left) and 2-beam

(right) direction at 2K. Bottom row: Ballistic simulations (no bulk scattering) for the geometries used in the measurement in the top row, with the boundary specularity set to p=0.1.

C) TEF spectra for various nozzle separations, scaled such that their first peaks coincide. Insert: The height of the primary focusing peak as a function of nozzle separation, from which the mean free path can be extracted to be about **λ** ~ 15 μm.

D) Temperature dependence of the TEF peaks for a nozzle pair with 2 μm separation. Insert: Peak height of the primary peak scaled by its value at 2 K as a function of temperature. The decay of the peak follows the reduction in mean-free-path.

## Figure 4: Specularity of FIB defined boundaries

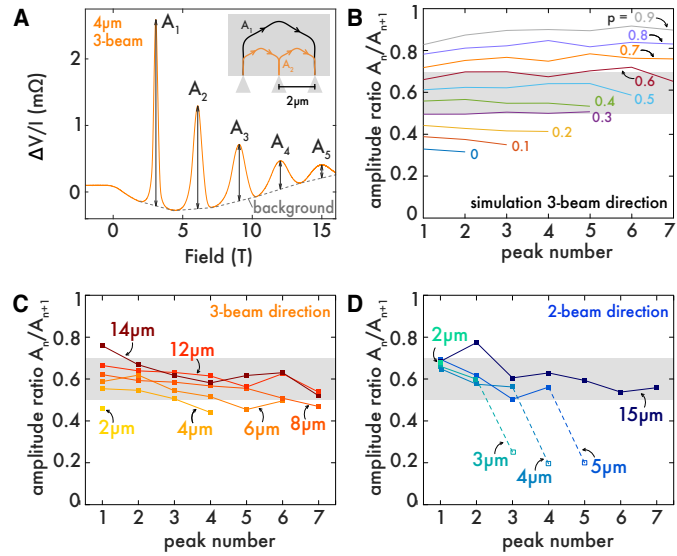

A) Definition of the peak amplitudes after subtraction of a smooth background which is obtained by connecting the voltage minima. The inset shows the trajectories giving rise to the first and second focusing peaks with amplitudes $A_1$ and $A_2$. Curiously, although the second focusing peak directly "bounces" off the boundary where there is another nozzle, its amplitude $A_2$ is not affected.

B) Simulated amplitude ratios for a range of boundary specularity values p between 0 and 1 in a purely ballistic model without bulk scattering. For small p values, the amplitude ratio q is strongly enhanced due to super-geometric focusing.

C) Amplitude ratio $q = \frac{A_n}{A_{n+1}}$ for various nozzle pairs measured along the 3-beam direction. The q-factor remains at a constant value of roughly $q = 0.6 \pm 0.1$ for all higher harmonic focusing events.

D) Along the 2-beam direction the extracted q-factor takes a similar value of $q = 0.6 \pm 0.1$ as in the 3-beam direction. When the cyclotron diameter becomes smaller than the nozzle width $b \approx 0.3\mu m$ at fields greater than $B = \frac{\hbar k_F}{2eb} \approx 11.3\,T$ the peak amplitude rapidly decreases in size (dashed lines, empty squares) and deviates from a power law behavior.

# Methods

## S1. Synthesis and characterization of PdCoO$_2$ crystals

Single crystals were grown in an evacuated quartz ampule with a mixture of PdCl$_2$ and CoO by the following methathetical reaction[33]: PdCl$_2$ + 2CoO → 2PdCoO$_2$ + CoCl$_2$. The ampule was heated at 1000 °C for 12 hours and stayed at 700-750 °C for 5 days. In order to remove CoCl$_2$, the resultant product was washed with distilled water and ethanol.

The orientation of the crystallographic axes was determined using the back-reflection Laue method. It was consistently found in over 5 crystals, that the in-plane a- and b-axes are rotated 90 degrees with respect to the hexagonal growth edges (see Fig. 2a). The out-of-plane c-axis lies perpendicular to the crystal platelets.

The residual resistance ratios extracted from the data in Fig. 2d are 457, 459 and 355 for $V_B$, $V_C$ and $V_D$ respectively. Although the size of the overall device exceeds the mean free path, ballistic effects at low temperatures may lead to a correction of the measured non-local voltage.

Quantitative energy dispersive X-ray spectroscopy (EDS) was used to confirm the elemental composition of the delafossite crystals using the AZtec software platform from Oxford Instruments. Typically, the oxygen concentration is severely underestimated due to a wrong carbon coating thickness, since carbon has an absorption edge near oxygen and heavily absorbs oxygen x-rays. Therefore, if the oxygen is fixed by stoichiometry to 2 ions, Pd and Co are found in equal atomic concentration.

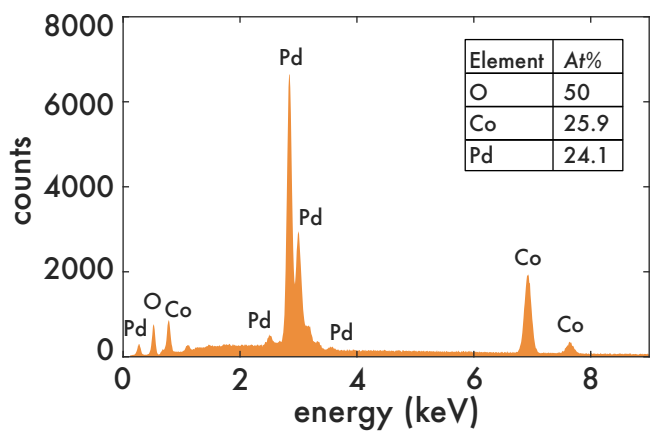

Figure S1: SEM-EDS spectrum of a PdCoO$_2$ crystals. Insert: Elemental analysis report after fixing the oxygen content by stoichiometry to 2.

## S2. Focused Ion Beam fabrication of point-like injections nozzles

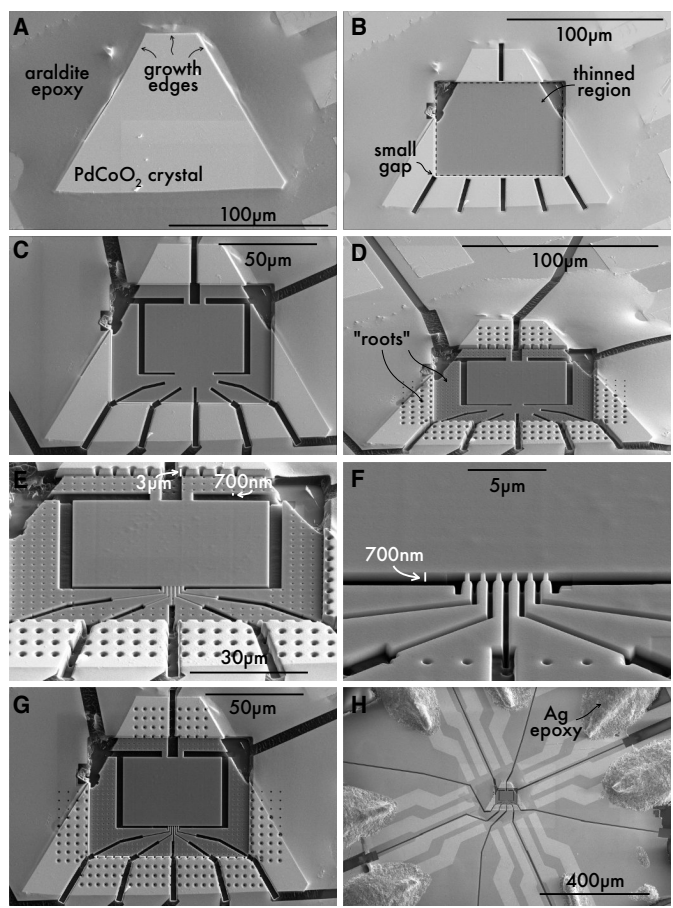

Figure S2: Step-by-step overview of the fabrication of a ballistic delafossite device using a Ga-based FIB.

A) A PdCoO$_2$ crystal is fixed onto a sapphire substrate (1.6×1.6×0.4 mm$^3$) with 5 min araldite® rapid epoxy, which is cured for 1 hour at 100°C on a hot plate. Care is taken to select a crystal with as little step edges as possible and with well-defined hexagonal growth edges, such that the crystallographic orientation can be directly determined.

B) The crystal is thinned down in the center to a final thickness of less than 1 μm (here 700 nm), using Ga$^{2+}$ ions at 30 kV, cutting a rectangle pattern with a current of 65 nA, 1 μs dwell time and the "dynamic all directions" scan option. Thinning down the crystal is a necessary step to fabricate narrow, closely spaced nozzles later on. Further ~10 μm wide rectangular cuts are made using 65 nA through the remaining thick parts of the crystal to define current and voltage contacts. A small gap is left to reduce re-deposition in the central area.

C) Rectangular cuts are patterned with 2.5 – 9.3 nA in the central region, which define a rectangular measurement region. The sides of the rectangle are polished with 2.5 nA under an angle of +1 degree with respect to the normal milling direction to obtain flat boundaries.

D) In order to ensure a homogenous current flow between all palladium layers despite having a top current injection, holes are patterned through the entire depth of the crystal with a current of 47nA and 2ms dwell time. At the inner edge of these 'root'-like features, the amorphous FIB-damage layer and re-deposition couples the individual layers together and increases the interlayer conductance. Roots are also milled into the central part of the device using a current of 2.5nA and 2ms dwell time.

E) The constrictions leading up to the nozzles are patterned with 80pA. Making long and thin constrictions is favorable, as they act as long flexures and reduce mechanical cracking of then nozzles due to strain from differential thermal contraction while cooling down.

F) The nozzles are cut using an array of cleaning cross section (CCS) cuts at 40pA, cut under an angle of 1degree. Initially the nozzles to a width of about 500nm and are then sequentially thinned down with CCSs until the final width of the nozzle is achieved.

G) Overview of the final device. If the nozzles are thinner than 350nm, a second layer of 5min araldite epoxy is added on top of the finished device and cured at room temperature for 24 hours. This eliminates nozzle fracture.

H) Final device on sapphire substrate. Silver wires were attached using Epotek EE129-4 silver epoxy and cured at 100°C for 1 hour. A 100 nm thick layer of sputtered Au connects the pre-evaporated gold leads on the substrate with the crystal device.

## S3. Long range Focusing

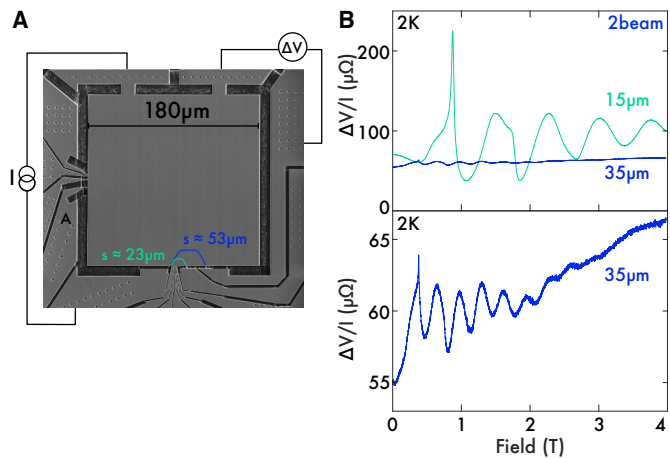

Figure S3: TEF across distances greater than the mean-free-path.

A) Measurement setup and propagation path through the sample. Along the 2-beam direction a nozzle separation of $L_1=15\,\mu m$ and $L_2=35\,\mu m$ corresponds to a path length ($s = \frac{3}{2}L$ in a perfect hexagon) through the device of $s_1 \approx 23\,\mu m$ and $s_2 \approx 53\,\mu m$ respectively.

B) Measured voltage $\Delta V$ divided by the applied current I = 6 mA as a function of transverse magnetic field at a temperature of 1.8 K. Top: Direct comparison of the signals of a 15μm and

35 µm separated nozzle pair. Bottom: Magnified signal of the 35 µm spaced nozzle pair. The double peak feature as well as 7 higher harmonic peaks are detectable.

## S4. Peak shape analysis from ballistic simulations

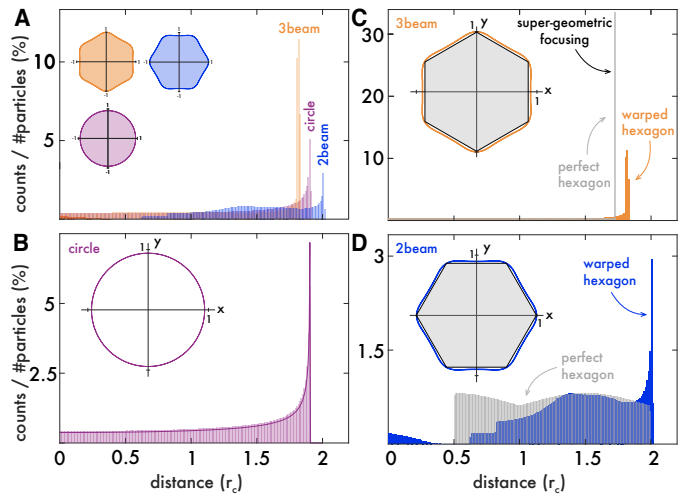

Figure S4: Comparison of the peak shapes for a circular, perfect and nearly perfectly hexagons

A) Comparison of the focusing spectra and peak heights of TEF in a circular FS and measured along the 3-beam and 2-beam direction in PdCoO$_2$. The cyclotron diameter is a factor $\frac{\sqrt{3}}{2}$ smaller in the 3beam direction compared to the 2beam direction, corresponding to the difference between inradius and circumradius of a hexagon. The diameter of the circular FS is expressed relative to the circumradius of the hexagon and was chosen smaller for clarity. Compared to a circular FS the 3beam direction has an enhanced peak, while the main peak along the 2beam direction is reduced and has a second broad hump.

B) In a circular FS the simulated focusing spectrum (shaded purple) diverges. For the mathematical derivation c.f. methods S5.

C) Comparison of the TEF spectra of a perfect hexagon and a hexagon with warped sides and rounded corners inferred from the FS of PdCoO$_2$ along the 3-beam direction. The perfectly flat edges lead to a geometrical enhancement of height of the focusing peak (so called super-geometric focusing or 'sfocusing'). In a warped hexagon, super-geometric focusing still enhances leads to an increased TEF peak.

D) Along the 2-beam direction, a perfect hexagonal FS does not exhibit any TEF at all, because there are not FS regions that are parallel to the nozzle injection direction. For a hexagon with rounded corners a focusing peak is recovered, analogous to the case of a circular FS.

## S5. Amplitude ratio analysis

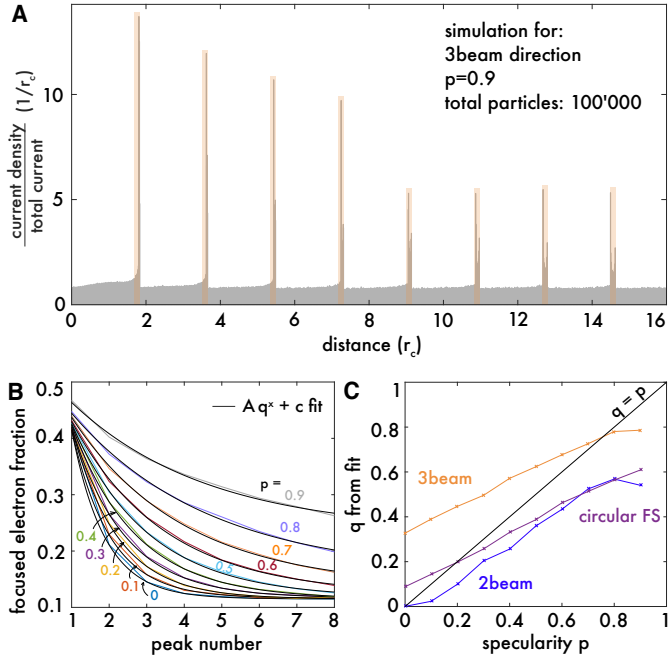

Figure S5: Simulations and analysis of TEF peaks for various boundary specularity coefficients.

A) Simulated TEF spectrum along the 3beam direction for a total of 100'000 particles with an isotropic incident angle distribution. The specularity of the boundary is p = 0.9, meaning that 90% of the electrons are specularly reflected upon impact with the boundary and the remaining 10% are assigned a random angle. During the simulation, all points of impact with the boundary for all particles are saved and displayed in the histogram above. For further analysis, the number of impacts in the orange shaded regions, corresponding to the TEF peak areas, is extracted.

B) The counts in the orange shaded area in panel A as a function of TEF peak number are plotted in color for a wide range of simulated specularity coefficients p between 0 and 0.9. The black line is a fit of the form $Aq^x + c$ to the data, where A scales the overall amplitude, q is the extracted "experimental specularity coefficient" and c is a background offset.

C) The extracted experimental amplitude ratio q as a function of the specularity coefficient p for the 3-beam (orange) and 2-beam (blue) direction as well as a circular FS (purple). The black line indicates where q=p.

The main result of these simulations is that the simple assumption of q = p, identifying the true surface specularity p with the measured power law coefficient q is not strictly applicable, even for a circular Fermi surface. Hence the amplitude ratio $\frac{A_{n+1}}{A_n}$ is a good indication, but not a perfect measure of the specularity of the boundary. The physical reason for this is two-fold. In the case of fully specular reflection the peak width grows with the number of peaks and due to their convolution with a finite nozzle size the measured voltage decreases with increasing peak number. In the opposite limit, even in the case of completely diffusive boundary scattering (p=0), a large number of TEF peaks are expected to arise from sfocusing. Therefore a simple analysis will extract a significant q value for p=0. Indeed we find q≈0.33 for diffuse scattering (figure S5c).

This value has a simple physical interpretation. Due the 3 main directions of propagation, approximately 1/3 of the electrons will be scattered into the direction that will be focused again. This statistical mechanism will lead to an apparent specularity of the boundary despite a completely diffusive scattering process. This is an alternative formulation of the super-geometric focusing properties of PdCoO$_2$.

## S6. Extraction of the mean-free-path λ

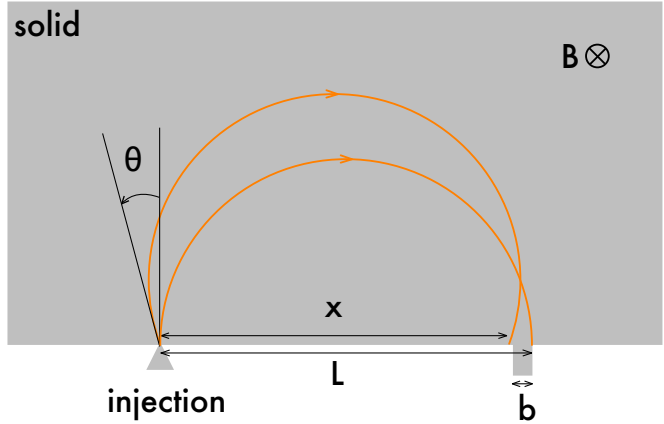

Figure S6: Geometrical model of TEF on a circular FS

The further two nozzles are spaced apart along the edge of the sample, the longer the path length s of an electron traveling through the bulk of the device, which increases the chances of being scattered away from its ballistic orbit. As pointed out by *Tsoi et al.*[31], the amplitude A$_1$ of the first TEF peak is proportional to $e^{-s/\lambda}$. The amplitude, however, also depends upon the ratio of $b/L$, where b is the width of the accepting nozzle and L is the distance between the nozzles. Assuming a point-like injection source and only the accepting nozzle having a width b, we find $x + b = L$, where L is the maximum distance an electron can travel at a fixed field ($L = 2r_c$). In a system with a circular FS, the travel distance x of an electron injected under an angle θ can be found by trigonometry to be $x = 2r_c \cos\theta$, where $r_c = \frac{\hbar k_F}{eB}$ denotes the cyclotron radius and $k_F$ is the Fermi momentum. With that and by Taylor expanding $\cos\theta \approx 1 - \frac{\theta^2}{2}$ for small θ, we find $\Delta\theta = 2\theta \approx 2\sqrt{\frac{2b}{L}}$. Accordingly, the amplitude of the first peak will decrease with increasing nozzle distance as $A_1 = 2\sqrt{\frac{2b}{L}} e^{-s/\lambda}$, where the path length is given by $s = \pi r_c$.

In the case of a hexagonal Fermi surface, the amplitude is similarly dependent on $e^{-s/\lambda}$ as well as the ratio $b/L$. For fitting the peak decay and extracting the mean-free-path λ in Fig. 3c we use the form $A_1(L) = A\ e^{-s/\lambda}\sqrt{\frac{b}{L}} + t$, with $b = 0.3\mu m$ and $s_{2beam} = \frac{3}{2}L$, $s_{3beam} = \sqrt{3}L$ are the path lengths for the 2-beam and 3-beam directions respectively. In addition to λ, the free variables are A, which sets the overall amplitude, and t, which takes the geometrical deviations from a non-circular FS into account. In the 3-beam direction the path-length is ill-defined due to the very nature of the super-geometric focusing effect. We choose the average between the longest and shortest path possible. The fit results are summarized in table 1. We note that this analysis is only valid for $\frac{b}{L} \ll 1$; once the nozzle width becomes comparable to the nozzle spacing the description breaks down. Further, particularly noticeable in the regime where $b \sim L$, but true in general, is that the maximum of the focusing peak does not occur at strictly $L = 2r_c$ but at lower magnetic fields where a nozzle of finite width can collect the maximum number of electrons.

|  | A [mΩ] | t [-] | λ [μm] |
| --- | --- | --- | --- |
| 2-beam direction | 7.7 | -0.0037 | 14 |
| 3-beam direction | 11.24 | 0.06 | 15.7 |

Table 1: Free parameters for fitting the peak decay of $A_1$ as a function of nozzle distance (c.f. Fig. 3c) with the form $A_1(L) = A\, e^{-s/\lambda} \sqrt{\frac{b}{L}} + t$. The small value of t in the 2-beam direction reflects the fact that the focusing in this orientation originates from the rounded corners of the hexagon which can be locally approximated by a circle. In the super-geometric focusing configuration the flat sides of the hexagon no longer resemble a circle leading to a larger t value.

### S7. Derivation of the TEF spectrum of a circular FS

In classical probability theory, let X and $\theta$ be continuous variables, where $X = g(\theta)$. The probability density function $f_\theta(\theta)$ describes the probability of $\theta$ falling within the infinitesimal interval [$\theta$, $\theta$+d$\theta$]. This can be transformed according to $f_X(x) = f_\theta(g^{-1}(x)) \cdot \left|\frac{d}{dx} g^{-1}(x)\right|$, which describes the probability of X falling into the interval [x, x+dx], in terms of the density if θ.

Let us consider the case of an (i) uncollimated and (ii) collimated beams of electrons injected into a TEF device. In all cases $x = g(\theta) = 2r_c \cos\theta$ is the travelling distance of electrons when injected at x=0.

(i) For an uncollimated beam $\theta$ has a uniform density on $[-\frac{\pi}{2}, \frac{\pi}{2}]$

$$f_\theta(\theta) = \begin{cases} \frac{1}{\pi} & \text{for } \theta \in \left[-\frac{\pi}{2}, \frac{\pi}{2}\right], \\ 0 & \text{otherwise.} \end{cases}$$

Requiring the probability density to be normalized, $\int_{-\infty}^{\infty} f_X(x)\, dx = 1$, we find:
$f_X(x) = \frac{2}{\pi} \frac{1}{\sqrt{4r_c^2 - x^2}}$, corresponding to the curve shown in figure S4b.

(ii) Similarly, for a beam which is collimated in a cosine form, we find its density

$$f_\theta(\theta) = \begin{cases} \frac{1}{2}\cos\theta & \text{for } \theta \in \left[-\frac{\pi}{2}, \frac{\pi}{2}\right], \\ 0 & \text{otherwise.} \end{cases}$$

The probability density function is then given by
$f_X(x) = \frac{x}{2r_c} \frac{1}{\sqrt{4r_c^2 - x^2}}$.

## S8. Numerical Methods

We start with a tight binding approximation of the FS[12] based on ARPES data[17],

$$k_F(\theta) = k_0 + k_6 \cos(6\theta) + k_{12} \cos(12\theta)$$

where $k_0 = 0.95 Å^{-1}$, $k_6 = 0.05\, Å^{-1}$, and $k_0 = 0.006\, Å^{-1}$. The equations of motion for an electron in an out-of-plane magnetic field $\boldsymbol{B} = B\hat{z}$ are

$$\hbar v = \frac{\partial \varepsilon}{\partial k}, \qquad \hbar \dot{\boldsymbol{k}} = -e\boldsymbol{E} + eB\hat{z} \times \boldsymbol{v}$$

where $\hbar$ is the reduced Planck's constant, $e$ is the charge of an electron, $\boldsymbol{v}$ is the Fermi velocity, and $E$ is the electric field experienced by the electron. In the ballistic regime, there is negligible electric field in the bulk, therefore we assume that $E = 0$. As discussed in the main text, the real space trajectory is a 90° rotation of the FS scaled by a factor of $\hbar/eB$. Because we are not concerned with transit times of the electrons, we can ignore the Fermi velocity $v$.

When interacting with an edge of the device, the probability of injecting into a particular state $n$ of the discretized Fermi surface is

$$p(n) = cos(\theta(n) - \phi)$$

where $\theta(n) = tan(v_y/v_x)$ is the direction of propagation of the state $n$ and $\phi$ is the angle of the normal to the edge. The Fermi surface is numerically discretized into states separated by constant arclength to remove the probability distribution's dependence on Fermi velocity[1]. The nearly perfectly hexagonal Fermi surface of $PdCoO_2$ has approximately flat edges which cause a high density of states to be injected at fixed angles.

Charge carriers are injected into a simplified two-dimensional version of the $PdCoO_2$ TEF device, beginning at a random position along the injection ohmic contact in an allowed state of the discretized FS (Fig. S7). These carriers then follow their semi-classical path[1], ignoring bulk scattering, until interacting with either an edge or ohmic contact of the device. In the case of a non-ohmic edge, a carrier is scattered into a new randomly state chosen according to the probability distribution for that edge. To ensure detailed-balance, floating voltage leads absorb an incident carrier and subsequently, the carrier is reemitted at a random position along the lead in a randomly chosen allowed state for that edge. The voltage at a lead is given by

$$V \propto \frac{\phi_{contact}}{L_{contact}}$$

where $\phi_{contact}$ is the number flux of carriers through the contact and $L_{contact}$ is the length of the device perimeter contacted by the voltage lead.

Electrons propagate within the device until they are absorbed by a grounded ohmic contact. The simulations of Fig. 2B of the main text are comprised of 1001 magnetic field points, each consisting of 30000 charge carriers, where the voltage difference between a TEF and a reference voltage lead all normalized by the voltage at the injecting contact $V_{source}$ is plotted. We observe qualitatively similar magnetic field dependence between this simulated ratio and the measured resistance of the real device for both tested orientations of device geometry relative to the crystal axis.

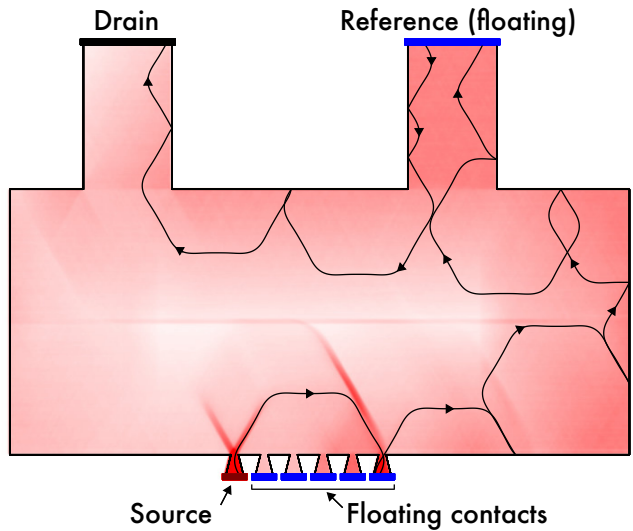

Figure S7: Monte Carlo simulation of ballistic trajectories

Normalized heat map of the position of electrons in the simulated TEF device geometry. Real space is divided into a grid. A count for each plaquette is incremented when an electron's trajectory passes through that plaquette. This count is reflected in the tone of red, with darker red corresponding to a higher count (where the count has been cut off at a high number to provide contrast in the bulk of the device). An example of such a trajectory is shown in black. Electrons are injected at the source (maroon contact) and are propagated until hitting the drain (black contact). Electrons incident on floating contacts (blue) are absorbed and reinjected at a random point along the contact. The Fermi surface can be freely rotated relative to the device to simulate both the 2- and 3- beam orientations.